\documentclass[pra,aps,twocolumn,superscriptaddress,showpacs]{revtex4}
\usepackage{graphicx,graphics,epsfig}
\usepackage{dcolumn}
\usepackage{bm}
\usepackage{amsmath}
\usepackage{verbatim}
\usepackage{color}
\usepackage[colorlinks=false]{hyperref} 
\usepackage{subfigure}
\usepackage{times,natbib}
\usepackage{amsmath,amsfonts,amssymb,graphics,graphicx,epsfig,color,times,natbib}
\usepackage{tikz}
\usepackage{verbatim}
\usepackage[standard]{ntheorem}

\date{\today}

\begin{document}
\title{State-independent contextuality sets for a qutrit}

\author{Zhen-Peng Xu}
 \affiliation{Theoretical Physics Division, Chern Institute of Mathematics, Nankai University,
 Tianjin 300071, People's Republic of China}

\author{Jing-Ling Chen}
\email{chenjl@nankai.edu.cn}
 \affiliation{Theoretical Physics Division, Chern Institute of Mathematics, Nankai University,
 Tianjin 300071, People's Republic of China}
 \affiliation{Centre for Quantum Technologies, National University of Singapore,
 3 Science Drive 2, Singapore 117543}

\author{Hong-Yi Su}
\email{hysu@mail.nankai.edu.cn}
 \affiliation{Theoretical Physics Division, Chern Institute of Mathematics, Nankai University,
 Tianjin 300071, People's Republic of China}

\begin{abstract}
We present a generalized set of complex rays for a qutrit in terms of parameter $q=e^{i2\pi/k}$, a $k$-th root of unity. Remarkably, when $k=2,3$, the set reduces to two well known state-independent contextuality (SIC) sets: the Yu-Oh set and the Bengtsson-Blanchfield-Cabello set. Based on the Ramanathan-Horodecki criterion and the violation of a noncontextuality inequality, we have proven that the sets with $k=3m$ and $k=4$ are SIC, while the set with $k=5$ is not. Our generalized set of rays will theoretically enrich the study of SIC proof, and experimentally stimulate the novel application to quantum information processing.
\end{abstract}

 \pacs{03.65.Ud,
03.67.Mn,
42.50.Xa}

\maketitle

\section{Introduction}
A three-dimensional quantum system (a qutrit) is the simplest system capable of exhibiting quantum contextuality. The first proof of state-independent contextuality (SIC) was given by Kochen and Specker (KS) using 117 rays in 1967, now renowned as the KS theorem~\cite{KS}. The KS theorem rules out noncontextual hidden variable models and demonstrates that quantum mechanics is essentially contextual. The rays form a set to test contextualtiy in a state-independent way. Later on Peres reported a 33-ray proof~\cite{Peres1991}, Conway and Kochen reported a 31-ray proof~\cite{ck1993},  followed by Yu and Oh~\cite{Yu-Oh} who  reduced the required number of observables to a record-breaking 13 rays and gave the state-independent proof of the KS theorem. In three dimensions, it has been proven that 13 is the minimal number of observables by Cabello~\cite{minimal}.


In general, a ray in the three-dimensional Hilbert space is described by a complex vector
\begin{eqnarray}
|v\rangle=(\sin\theta\cos\phi e^{i\tau_0},\sin\theta\sin\phi e^{i\tau_1},\cos\theta e^{i\tau_2}),
\end{eqnarray}
where $\theta,\phi,\tau_i$ are real parameters.
The rays are usually taken as real-valued, i.e., all phase factors vanish (e.g., see the literature mentioned above), since they can be conveniently viewed as vectors in the three-dimensional Euclid space and applied to practical experiments.
Very recently, Bengtsson, Blanchfield, and Cabello (BBC) suggested that it would have advantages if the rays are taken as complex vectors, and they proposed 21 complex vectors to detect SIC~\cite{Cabello6}.

In this paper, we present a set of $n$ rays (unnormalized), which we recast into three classes as follows:
\begin{subequations}\label{ray}
\begin{align}
({\rm i}):&\;\;\;(1,0,0), (0,1,0), (0,0,1),\\
({\rm ii}):&\;\;\;(1,-q^i,0), (1,0,-q^i), (0,1,-q^i),\\
({\rm iii}):&\;\;\;(1,q^i,q^j),
\end{align}
\end{subequations}
where $q=e^{2\pi i/k}$ is a $k$-th root of unity, and $i,j=1,2,\ldots, k$. In general, the first class contains $3$ vectors, the second contains $3k$, and the third contains $k^2$, hence the total $n=3+3k+k^2$ vectors. By a direct calculation, let $k=2$, then from Eq.~(\ref{ray}) one obtains the Yu-Oh 13 rays:
\begin{subequations}
\begin{align}
&(1,0,0), (0,1,0), (0,0,1),\\
&(1,1,0), (1,0,1), (0,1,1),\nonumber\\
&(1,-1,0), (1,0,-1), (0,1,-1),\\
&(1,1,1), (1,1,-1),(1,-1,1),(1,-1,-1),
\end{align}
\end{subequations}
and let $k=3$, one obtains 21 vectors equivalent to those in BBC's proposal.

Remarkably, the Yu-Oh SIC set and the BBC SIC set share a unified structure as in Eq.~(\ref{ray}). This raises a natural question: What if $k$ is arbitrary? In this paper, we shall address this question. The paper is organized as follows. In Sec.~II, we shall study the problem by the Ramanathan-Horodecki (RH) criterion~\cite{Horo}. Explicitly, we calculate the fractional chromatic number $\chi_f$ up to $k=5$. We find that for $k=5$, the set is not SIC, and that for $k=4$, we can construct an explicit inequality to identify SIC. In Sec.~III, we shall use the inequality criterion to study the problem. To be specific, we calculate the quantum violation of the maximally mixed state (MMS) and find that when $k=3m$ the sets are SIC. In the last section, we make a discussion and conclusion.

\section{RH criterion to SIC}
Given a set of rays, they can be represented as an exclusivity graph $G$ whose vertices stand for rays and two vertices are connected if and only if the corresponding two rays are orthogonal~\cite{CSW10,CSW14}. Recently Ramanathan and Horodecki proposed a necessary condition~\cite{private} for a set of rays to be SIC for dimension $d$
\begin{eqnarray}
\chi_f(G)>d,
\end{eqnarray}
with $\chi_f(G)$ being the fractional chromatic number of the exclusivity graph~\cite{Scheinerman}. To make the paper self-contained, we briefly review the chromatic number $\chi(G)$ and fractional chromatic number $\chi_f(G)$.

\begin{definition}
An $\ell$-coloring of a graph $G$ is an assignment of $\ell$ colors to vertices so that adjacent vertices receive different colors. The chromatic number of $G$, denoted $\chi(G)$, is the least $\ell$ for which $G$ has an $\ell$-coloring.
\end{definition}

\begin{definition}
A $a:b$-coloring of a graph $G$ is an assignment of sets of size $b$ out of $a$ available colors to vertices of a graph such that adjacent vertices receive disjoint sets. The $b$-fold chromatic number $\chi_b(G)$ is the least $a$ such that an $a:b$-coloring exists. The fractional chromatic number $\chi_f(G)$ is defined to be
\begin{eqnarray}
\chi_{f}(G) = \inf_{b}\frac{\chi_{b}(G)}{b}.
\end{eqnarray}
Note that $\chi_{f}(G)$ can be obtained at a finite $b$.
\end{definition}

\begin{figure}[h]
\centering
\includegraphics[width=0.45\textwidth]{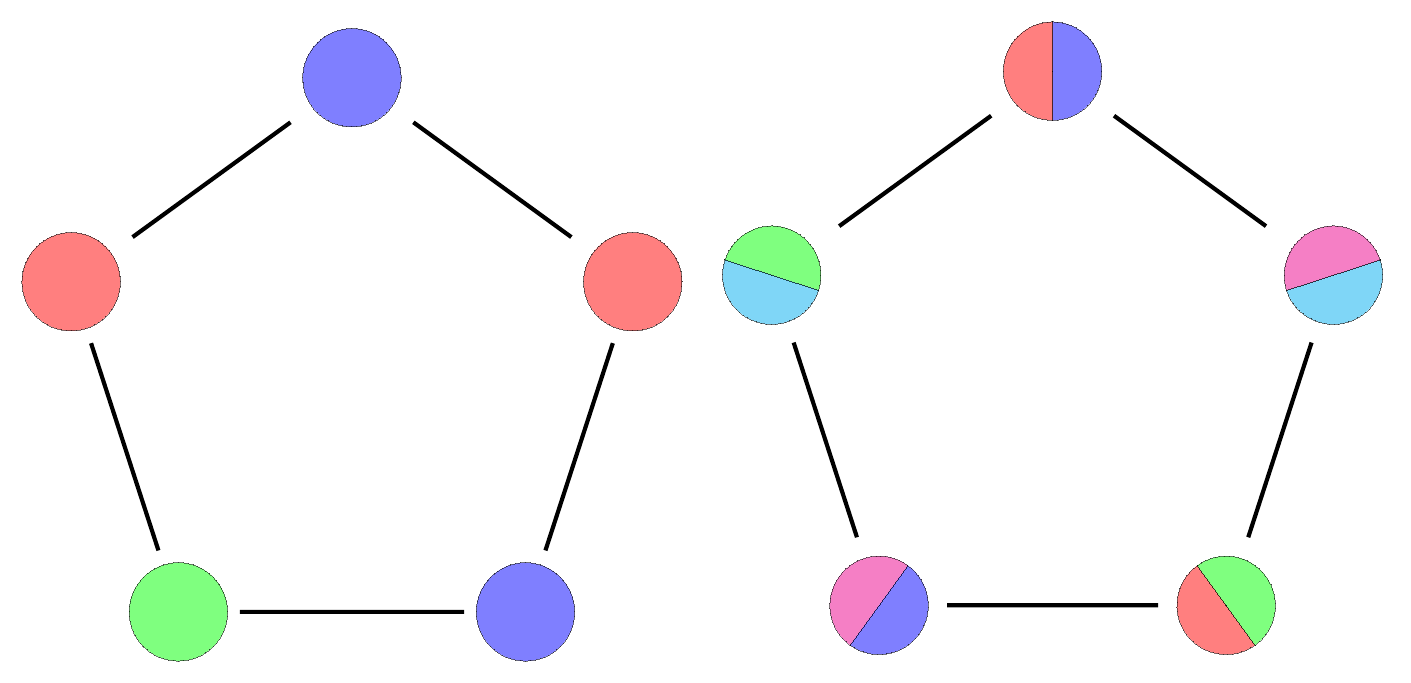}
\caption{(Color online) The coloring (left) and the $5:2$-coloring (right) for pentagon.}
\label{pentagon}
\end{figure}

As an example, let us consider a pentagon graph, for which $\chi(G)=3,\;\chi_f(G)=5/2$ (see Fig.~\ref{pentagon}).

For Eq.~(\ref{ray}), there are $n=3+3k+k^2$
rays for a contextual test. Such rays form an exclusivity graph with $n$ vertices through the orthogonal relations of these vectors. The key point to use the RH criterion is to calculate $\chi_f(G)$. If $\chi_f(G)\leq d$, then the set is not SIC.  In Table I, we have computed $\chi_f$ up to $k=5$ within our strength. Based on the RH criterion, one can confirm the set with $k=5$ is not SIC.
\begin{table}[h]
\centering
\begin{tabular}{lllll}
\hline\hline
  $k$      & 2 & 3 & 4 & 5 \\ \hline
    $n$      & 13 & 21 & 31 & 43 \\ \hline
$\chi_f$   & $35/11$ & $10/3$ & $67/21$   &  3     \\
\hline\hline
\end{tabular}
\caption{$\chi_f$ for $k=2,3,4,5$.}
\end{table}

It is well known that the sets with $k = 2,3$ are SIC. The following noncontextuality inequality shows that it is also true when $k=4$:
\begin{eqnarray}
5 \sum_{i \in C_1} \hat P_i + 3 \sum_{i \in C_2} \hat P_i + \sum_{i \in C_3} \hat P_i \le 21,\label{k4}
\end{eqnarray}
where $C_1$ means the rays in the first class in Eq.~(\ref{ray}) and so on. Quantum mechanically for any state, the left-hand side of (\ref{k4}) equals $67/3$, larger than the classical bound 21.
%

\section{Inequality criterion to SIC}
For $k\geq6$, we resort to a special form of noncontextuality inequality to investigate SIC. The inequality we consider here is given by
\begin{eqnarray}
\sum_{i=1}^n \hat P_i\leq\alpha,\label{inequality}
\end{eqnarray}
where $\hat P_i=|v_i\rangle\langle v_i|$, $|v_i\rangle$'s are the rays in the set,  and $\alpha$ is the classical bound which is equal to the independence of a graph $G$.

Directly computation shows that
\[
\sum_{i=1}^n \hat P_i = (1 + k + k^2/3)\openone = (n/3)\openone.
\]
So, the set is a SIC one if $\alpha < n/3$.

For convenience, we say the independence of the rays means the independence of the graph $G$ which represents these rays, and two rays are independent means they are not orthogonal.
Indeed, $\alpha$ is the maximal number of mutually nonorthogonal vectors in a set. It can be obtained  that
\begin{eqnarray}\label{alpha}
\alpha= \begin{cases}
\frac{k^2}{3}+k, & {\rm for\;} k=3m, \\
1+k^2, &{\rm for\;} k\neq3m,
\end{cases}
\end{eqnarray}
so that
\begin{eqnarray}
\begin{cases}
\alpha < n/3, & {\rm for\;} k=3m, \\
\alpha > n/3, &{\rm for\;} k\neq3m.
\end{cases}
\end{eqnarray}
Therefore, (\ref{ray}) with $k=3m$ is an SIC set, which can be detected by inequality (\ref{inequality}). This is confirmed by numerical results in Table II.
\begin{table}[h]
\centering
\begin{tabular}{llllllll}
\hline\hline
  $k$      & 6 &7&8& 9&10&11 & 12 \\ \hline
    $n$      & 57 &73&91& 111&133&157 & 183 \\ \hline
      $\alpha$      & 18 &50&65& 36&101&122 & 60 \\ \hline
$n/3$ & $19$ & 73/3 & 91/3 & $37$ &133/3 & 157/3 & $61$   \\ 
\hline\hline
\end{tabular}
\caption{$\alpha$ and $n$ for $k=6,7,...,12$.}
\end{table}

\begin{figure*}[t]
\centering
\includegraphics[width=0.8\textwidth]{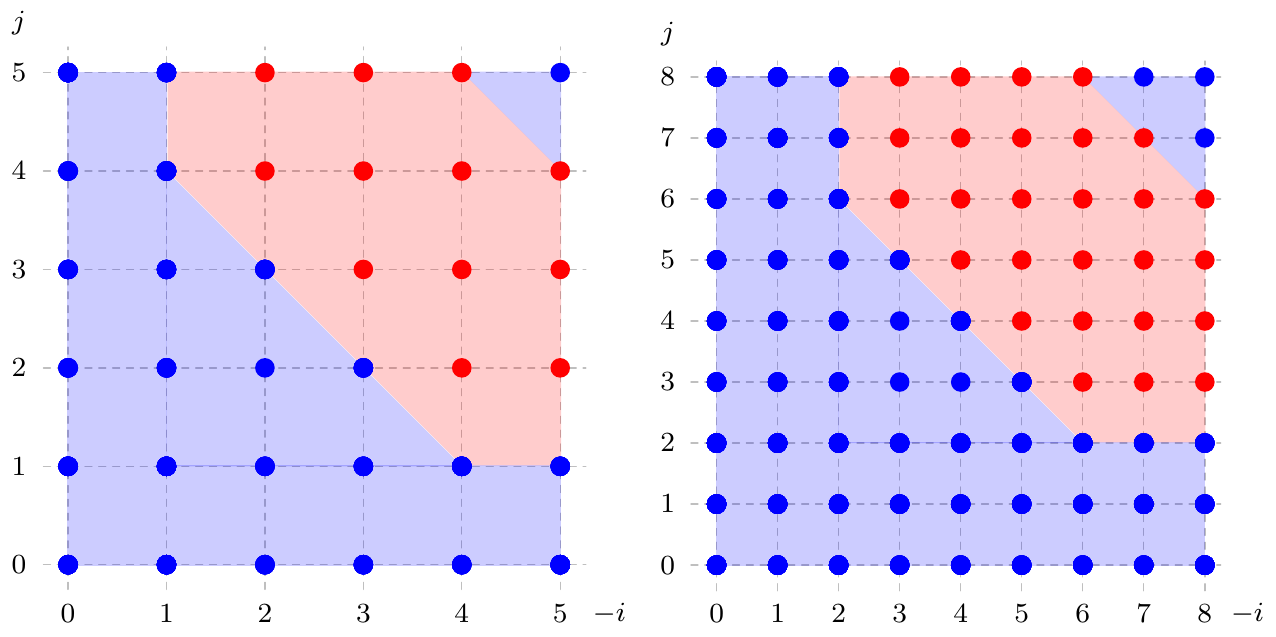}
\caption{(Color online) The rays chosen (red points) in the third class for $k=6$ (left) and $9$ (right).}
\label{Fig2}
\end{figure*}

In the following, we shall analyze how to obtain Eq.~(\ref{alpha}).

\emph{Case 1}: $k\neq 3m$. In this case, we have

\noindent(a) vectors in the first class in Eq.~(\ref{ray}) are mutually orthogonal;

\noindent(b) vectors in the third class in Eq.~(\ref{ray}) are mutually nonorthogonal;

\noindent(c) any vector in the first class will be nonorthogonal with any vector in the third class;

\noindent(d) any vector in the second class will be orthogonal with some vectors in the third class.

\noindent Hence, $\alpha=1+k^2$ (one ray from the first class plus all rays from the third class).

\emph{Case 2}: $k=3m$. Due to the algebraic relation
\begin{eqnarray}
1+e^{i2\pi/3}+e^{i4\pi/3}=0,
\end{eqnarray}
vectors in the third class could be orthogonal. This is different from \emph{Case 1}.
It is easy to see that  the solution to
\begin{eqnarray}
1+q^i+q^j = 0,
\end{eqnarray}
is
\begin{eqnarray}
i=-j = m~\text{or}~-m.
\end{eqnarray}

Two rays $(1,q^{i_1},q^{j_1}), (1,q^{i_2},q^{j_2})$ are orthogonal if and only if $1+q^{i_2-i_1}+q^{j_2-j_1} = 0$. We have
\begin{eqnarray}
 i_2 = {\rm mod}(i_1+m, 3m), j_2 = {\rm mod}(j_1-m, 3m),
\end{eqnarray}
or
\begin{eqnarray}
 i_2 = {\rm mod}(i_1-m, 3m), j_2 = {\rm mod}(j_1+m, 3m).
\end{eqnarray}
So,  $(1,q^{i},q^{j}), (1,q^{i+m},q^{j-m}), (1,q^{i-m},q^{j+m})$ form a complete orthogonal basis for each $i,j$. Thus only one of them should appear in the independent set. Totally, there are $3m^2 = k^2/3$ rays can be chosen at most in the third class.


If $(1,-q^i,0)$ is chosen in the second class, then $(1,q^{i},q^{j})$ cannot be chosen for every $j$; If $(1,0,-q^j)$ is chosen in the second class, then $(1,q^{i},q^{j})$ can not be chosen for every $i$; If $(0,1,-q^\mu)$ is chosen in the second class, then $(1,q^{i},q^{j})$ can not be chosen for every $j-i=\mu$. To find the independence of the vectors in the second and third classes, one can choose $k$ rays from the second class and $k^2/3$ from the third class (in Fig.~\ref{Fig2} we have plotted how to choose rays in the third class for $k=6,9$), hence $k+k^2/3$ rays in total. Explicitly, the maximal independent set of $k+k^2/3$ rays are as follows:
\begin{itemize}
\item choose none from the first class;
\item choose $(1,-q^i,0), (1,0,-q^j),(0,1,-q^\mu)$ for ${\rm mod}(-i,3m),\;j,\;{\rm mod}(-\mu-1,3m)=0,\ldots,m-1$;
\item choose $(1,q^i,q^j)$ for ${\rm mod}(-i,3m),\;j,\;{\rm mod}(i-j-1,3m) \neq 0,\ldots,m-1$.
\end{itemize}
This claims Eq.~(\ref{alpha}).

\section{Conclusion and discussion}

In summary, we have presented a generalized set of rays in terms of parameter $q=e^{i2\pi/k}$ (cf. Eq.~(\ref{ray})) for a qutrit. When $k=2,3$, the set corresponds to the Yu-Oh set and the BBC set, respectively. Based on the RH criterion and the violation of inequality (\ref{inequality}), we have proven that the sets with $k=3m$ and $k=4$ are SIC, while the set with $k=5$ is not. Thus our generalized set of rays will enrich the study of SIC proof, and stimulate the new application to quantum information processing~\cite{QIP}.

Let us make a discussion to end this paper. In comparison to inequality (\ref{inequality}), the more general inequality is given by
\begin{eqnarray}
\sum_{i=1}^n w_i \hat P_i\leq\alpha,\label{inequality-w}
\end{eqnarray}
where $w_i$ are weights. For simplicity, in this paper we have only considered the case where all weights are unity.
By adjusting the weights, it is possible to identify the set with $k\neq3m$ is SIC, if (\ref{inequality-w}) is violated. We shall investigate this subsequently.


\begin{acknowledgments}
We thank A. Cabello for useful discussions and for bringing Ref.~\cite{Cabello6} to our attention. J.L.C. is supported by the National Basic Research Program (973
Program) of China under Grant No.\ 2012CB921900 and the NSF of China
(Grant Nos.\ 11175089 and 11475089). This
work is also partly supported by the National Research Foundation
and the Ministry of Education, Singapore.
\end{acknowledgments}


\begin{thebibliography}{99}
\bibitem{private}
Private correspondence with A. Cabello, who showed that the RH criterion is a necessary but not sufficient condition for SIC.
\end{thebibliography}
\end{document}